\documentclass[conference]{IEEEtran}
\IEEEoverridecommandlockouts
\usepackage{cite}
\usepackage{amsmath,amssymb,amsfonts}
\usepackage{algorithmic}
\usepackage{graphicx}
\usepackage{textcomp}
\usepackage{xcolor}
\usepackage{acro}
\usepackage[caption=false,font=footnotesize]{subfig}
\usepackage{url}

\usepackage{cleveref}
\crefname{figure}{Fig.}{Figs.}
\Crefname{figure}{Figure}{Figures}
\crefname{table}{Table}{Tables}
\Crefname{table}{Table}{Tables}
\crefname{equation}{}{}
\Crefname{equation}{Equation}{Equations}

\DeclareAcronym{llm}{
  short = LLM,
  long  = large language model,
}

\DeclareAcronym{vlm}{
  short = VLM,
  long  = vision language model,
}

\def\BibTeX{{\rm B\kern-.05em{\sc i\kern-.025em b}\kern-.08em
    T\kern-.1667em\lower.7ex\hbox{E}\kern-.125emX}}
\begin{document}

\title{PolyMemDB: A Polyglot Database System for AI Memory Management}

\author{\IEEEauthorblockN{Yu Wang}
\IEEEauthorblockA{\textit{University of Helsinki} \\
yu.wang@helsinki.fi}
\and
\IEEEauthorblockN{Jiaheng Lu}
\IEEEauthorblockA{\textit{University of Helsinki} \\
jiaheng.lu@helsinki.fi}}

\maketitle

\begin{abstract}
With the widespread adoption of personal intelligent agents, users generate massive, heterogeneous data during long-term interactions. Leveraging this data as long-term memory helps reduce token overhead and deliver personalized experiences. However, existing memory systems face two primary limitations: they rely on single-storage paradigms that fragment multi-dimensional data, and they lack fine-grained data provenance to resolve long-term factual conflicts, thereby worsening \ac{llm} hallucinations.

In this demonstration, we introduce PolyMemDB, a novel system tailored for managing agent memory. PolyMemDB has a polyglot storage architecture designed to track and manage various memory data types, including graph, vector, probability and spatial-temporal data. To ensure factual consistency and reduce hallucinations, it features a probabilistic inference engine that integrates temporal decay with semiring aggregation, resolving long-term factual conflicts, providing detailed data provenance, and enabling users to trace reasoning chains transparently.
\end{abstract}

% \begin{IEEEkeywords}
% Agent memory, Polyglot storage, Data provenance
% \end{IEEEkeywords}

\section{introduction}
\subsection{Motivation and Contributions}

\begin{figure*}[htbp]
    \centering
    \includegraphics[width=0.85\textwidth]{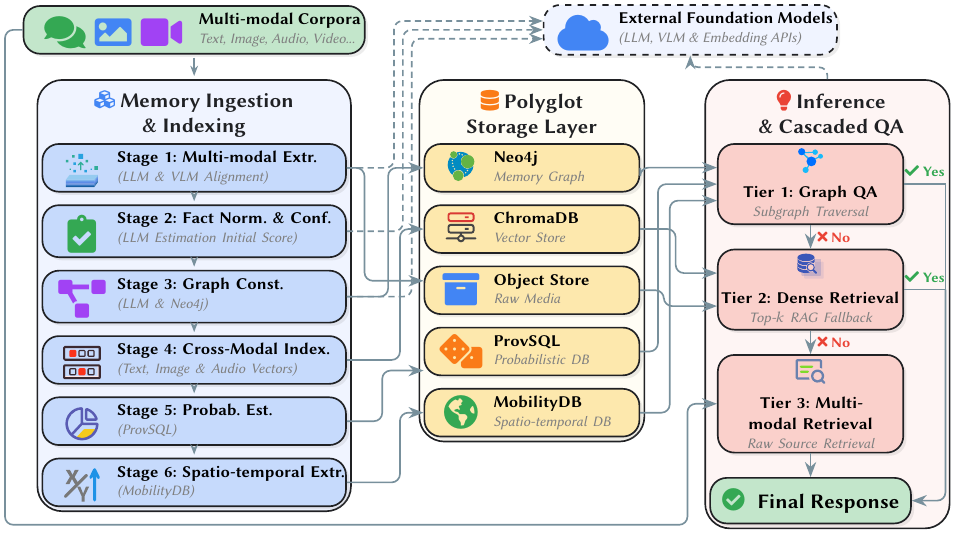} 
    \caption{PolyMemDB architecture: multi-modal ingestion, polyglot storage, and cascaded QA routing.}
    \label{fig:system_arch}
\end{figure*}

The rapid advancement of agent technology based on \acp{llm} has enabled the creation of personal assistants capable of managing long-term interactive tasks. In this context, the memory mechanism has emerged as a key factor influencing both system performance and user experience~\cite{zhangSurveyMemoryMechanism2025a}. By structuring and persistently storing the historical context generated during an agent's operation, it is possible not only to reduce inference latency and token costs through the reuse of semantically similar responses but also to accurately capture users' personalized preferences and enable the agent to abstract higher-order generalized knowledge from them~\cite{zhangSurveyMemoryMechanism2025a}.

However, in real-world scenarios, the management and maintenance of long-term memory face two core challenges: 1)  Long-term interactions generate massive amounts of heterogeneous context with diverse attributes (such as complex entity networks, logical relationships, and spatio-temporal constraints). Existing systems generally rely on a single storage paradigm, making it difficult to support efficient storage and retrieval of multimodal heterogeneous information; 2) During long-term memory evolution, systems typically employ coarse-grained state overwriting and lack factual maintenance mechanisms that integrate temporal sequences and probability, resulting in an urgent need to improve data provenance capabilities and the ability to address \ac{llm} hallucinations.

To tackle these challenges, we developed and implemented \textbf{PolyMemDB}, which combines multi-modal storage, graph databases, and spatio-temporal indexing technologies. It also features graph updates that integrate data provenance with temporal decay, facilitating conflict resolution for multi-source facts. In this demonstration, our core contributions are summarized as follows:

\begin{itemize}
\item \textbf{Polyglot storage layer with multiple databases:} We integrated various heterogeneous storage components, including multi-modal object storage, graph databases, spatio-temporal databases, and probabilistic databases. This architecture addresses the limitations of traditional text-only and vector-based retrieval.
\item \textbf{Probabilistic memory graph maintenance:} To address the challenges of event integration and factual conflicts in long-term interactions, we introduce a dynamic graph update mechanism that combines temporal decay and data provenance. By extending semiring inference in probabilistic databases to the evidence chain, the system can resolve conflicts and align entities, thereby mitigating \ac{llm} hallucinations while preserving interpretability.
\end{itemize}

\subsection{Comparison to Existing Systems}

Recent research has explored various aspects of structured representations and retrieval mechanisms for agent memory. For example, at the level of data organization, MemForest~\cite{chenMemForestEfficientAgent2026} approaches memory as a problem focused on efficient time-based writes, creating a tree structure that organizes time-related information and allows updates to occur only along specific paths within the tree. This method avoids the need to rewrite the entire global memory state. HyperMem~\cite{yueHyperMemHypergraphMemory2026} improves traditional graph databases by introducing a three-tier hypergraph model of topics, fragments, and facts, using hyperedges to capture complex dependencies among discrete memory fragments. Unlike MemForest, which focuses on text-time indexing, and HyperMem, which focuses on homogeneous hypergraph structures, PolyMemDB addresses the scheduling problem of heterogeneous multi-dimensional data. PolyMemDB abandons the single-storage paradigm and introduces a polyglot storage architecture at the lower level. It not only utilizes a graph database (Neo4j) to maintain entity topology but also pushes complex spatio-temporal constraints directly down to a spatio-temporal database (MobilityDB) for joint filtering of dynamic time windows and geographic polygons, overcoming the limitation of existing systems that can only process pure text relationships. 

At the level of conflict resolution mechanisms, MRAgent~\cite{jiMemoryReconstructedNot2026} constructs a ``Cue-Tag-Content'' graph and relies on an active reconstruction mechanism, enabling the \ac{llm} to iteratively explore and prune paths based on intermediate states during the retrieval phase. PolyMemDB extends the semiring inference in probabilistic databases. For memory triplets, PolyMemDB appends a historical observation sequence to the edge attributes and, through probabilistic inference and temporal decay, computes distribution vectors for positive drive, negative inhibition, and evidence conflict at the database level. This enables PolyMemDB to provide fine-grained data provenance while resolving cognitive conflicts through underlying mathematical inference, thereby enhancing the system’s reliability.

\begin{figure*}[htbp]
    \centering
    
    % 第一张子图
    \subfloat[Main dashboard.\label{fig:demo_main}]{%
        \begin{minipage}[t]{0.64\textwidth}
            \centering
            \includegraphics[width=\linewidth]{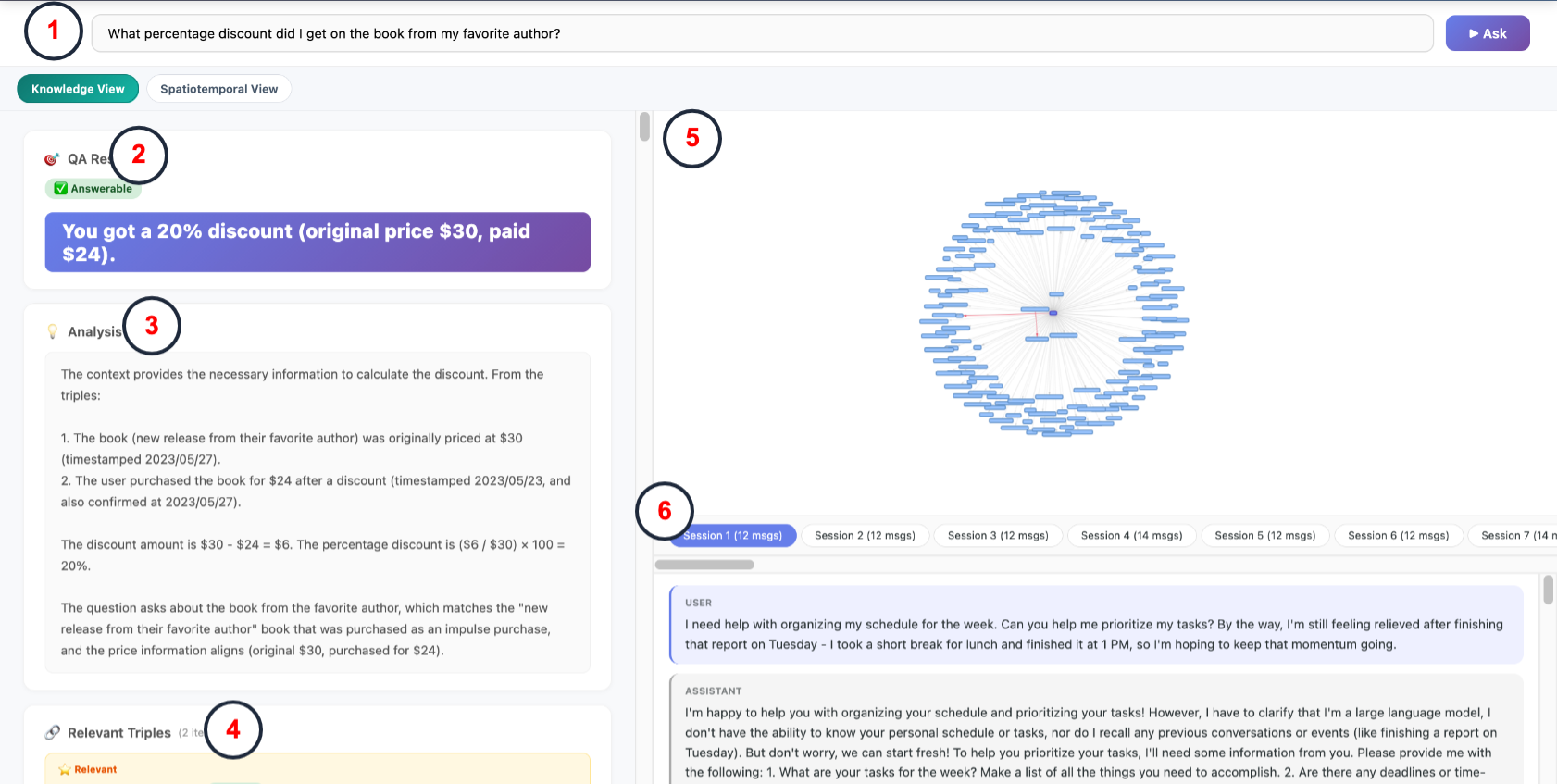}
        \end{minipage}%
    }
    \hfill
    \subfloat[Evidence panel.\label{fig:demo_triples}]{%
        \begin{minipage}[t]{0.34\textwidth}
            \centering
            \includegraphics[width=\linewidth]{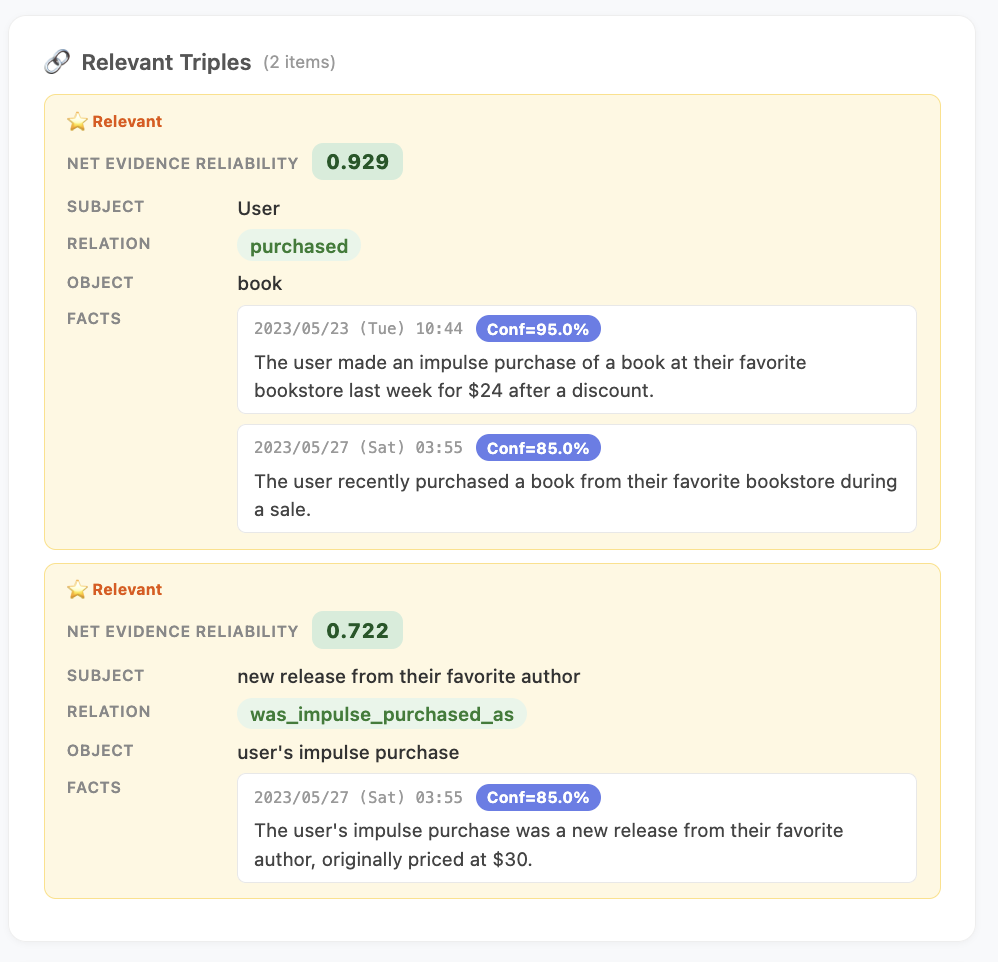}
        \end{minipage}%
    }
    
    \caption{PolyMemDB UI: (a) Main dashboard for query and graph visualization; (b) Evidence panel showing probabilistic facts.}
    \label{fig:system_demo}
\end{figure*}

\section{system overview}

As shown in~\cref{fig:system_arch}, the workflow of PolyMemDB comprises three core stages:

First, in the Memory Ingestion \& Indexing, the system decomposes the input data into atomic facts. During this process, PolyMemDB leverages \acp{llm} and \acp{vlm} to perform entity alignment, execute coreference resolution and ellipsis recovery, and complete named entity extraction, fact normalization, and confidence estimation.

Subsequently, it enters the Polyglot Storage Layer. To fully capture the complex associations in the data, PolyMemDB performs targeted extraction and routing of storage for normalized facts based on their multi-dimensional features. Specifically, the system constructs entities, attributes, and their evolutionary trajectories across long periods into a connected relationship graph (relying on Neo4j), providing structural support for subsequent complex chain inference; it stores event confidence in a probabilistic database (relying on ProvSQL); and parses complex spatio-temporal constraints for management by a spatio-temporal database (relying on MobilityDB). Meanwhile, embedding vectors and raw context content are persisted into a vector database (relying on ChromaDB) and object storage, respectively.

Finally, when responding to user interactions, the system triggers the Inference \& Cascaded QA. To balance inference overhead and recall accuracy, PolyMemDB designs a top-down, three-tier federated retrieval mechanism. The system first executes \textit{Tier 1 (Graph QA)}, prioritizing low-latency subgraph traversal and jointly utilizing the probabilistic database and spatio-temporal index for rule filtering and logical inference; if this tier yields no answer, it triggers a fallback mechanism to enter \textit{Tier 2 (Dense Retrieval)}, utilizing the query's embedding vectors to perform Top-$k$ semantic retrieval of structured facts within the vector database; if sufficient evidence is still not found, the system ultimately drills down to \textit{Tier 3 (multi-modal Retrieval)} to directly retrieve raw data snippets, thereby generating the final response.

\section{probabilistic memory graph maintenance}

Event consolidation and conflict resolution are core challenges in long-term memory management for agents. To this end, PolyMemDB designs a dynamic graph update mechanism that combines temporal decay and probabilistic inference.

PolyMemDB formalizes the agent's memory as a memory graph $\mathcal{G} = (V, E, \alpha)$. Here, $V$ is the set of heterogeneous entities (e.g., \textsc{Person}, \textsc{Concept}, etc.); $E \subseteq V \times \mathcal{L} \times V$ is the set of directed relationship edges; and $\alpha$ is the attribute labeling function. We define a single fact as a tuple $f_i = \langle \rho_i, c_i, t_i \rangle$. Here, $\rho_i \in \{+, -\}$ denotes the sentiment polarity of the evidence, $c_i \in [0, 1]$ is the initial confidence assigned by the \ac{llm}, and $t_i$ is the timestamp of the fact observation. Before inserting a fact into the memory graph, the system retrieves the Top-$k$ local subgraphs via semantic similarity retrieval as context, which the agent uses to autonomously decide whether to create or update nodes or edge relationships, thereby maximizing the accuracy of entity alignment.

 To effectively mitigate \ac{llm} hallucinations and preserve the interpretability of raw materials, PolyMemDB abandons the simple state overwriting mechanism. For any triplet $e = (u, l, v) \in E$ in the graph, the system appends and maintains a historical observation sequence $F = (f_1, f_2, \dots, f_T)$ (satisfying $t_1 \le t_2 \le \dots \le t_T$) in its edge attributes. This mechanism fully preserves the evidence observed at different time points.

 When calculating the confidence of a specific triplet, we extend the semiring framework found in probabilistic databases (e.g., ProvSQL~\cite{senellartProvSQLProvenanceProbability2018}). Since the influence of real-world events decays over time, we introduce an exponential decay factor $\lambda \in (0, 1]$. For a fact $f_i$ in the sequence with a temporal distance index of $\Delta_i = T - i$, its dynamic weight is calculated as $w_i = c_i \cdot \lambda^{T - i}$. Based on the inclusion-exclusion principle, we partition the evidence sequence $F$ into a positive set $F^+$ and a negative set $F^-$, and calculate their cumulative joint confidences, respectively, as defined in \cref{eq:cumulative_prob}:
\begin{equation}
\label{eq:cumulative_prob}
P^+ = 1 - \prod_{f_i \in F^+} (1 - w_i), \quad P^- = 1 - \prod_{f_j \in F^-} (1 - w_j)
\end{equation}

Based on $P^+$ and $P^-$ derived from \cref{eq:cumulative_prob}, PolyMemDB computes four intermediate cognitive states to capture the interactions between conflicting evidence: positive drive $S_{action} = P^+(1 - P^-)$, negative inhibition $S_{inaction} = P^-(1 - P^+)$, evidence conflict $S_{conflict} = P^+P^-$, and absence of evidence $S_{ignorance} = (1 - P^+)(1 - P^-)$.

To comprehensively evaluate the factual reliability, PolyMemDB aggregates these intermediate states into a \textit{Net Evidence Reliability} score, denoted as $\mathcal{R} \in [-1, 1]$:

\begin{equation}
\label{eq:net_evidence}
\mathcal{R} = (S_{action} - S_{inaction}) \cdot (1 - S_{conflict} - S_{ignorance}) 
\end{equation}

Here, $\mathcal{R}$ compactly integrates all evidence dimensions. The base term $(S_{action} - S_{inaction})$ establishes the directional truthfulness, yielding a positive score for supported facts and a negative score for refuted ones. The multiplier $(1 - S_{conflict} - S_{ignorance})$ serves as a dynamic penalty, reducing the net reliability $\mathcal{R}$ when there are significant factual disputes or insufficient evidence. When $\mathcal{R} \approx 0$, identifying the maximum intermediate state ($\arg\max$) clearly distinguishes whether the low score stems from contradictory facts ($S_{conflict}$) or an evidence void ($S_{ignorance}$). In the subsequent QA stage, providing $\mathcal{R}$ and these decomposed distributions as quantitative references empowers the \ac{llm} to avoid binary hallucinations and generate provenance-grounded answers.

\section{demonstration}

\begin{figure*}[htbp]
    \centering
    \subfloat[Knowledge view.\label{fig:italy_knowledge}]{%
        \begin{minipage}[t]{0.33\textwidth}
            \centering
            \includegraphics[width=\linewidth]{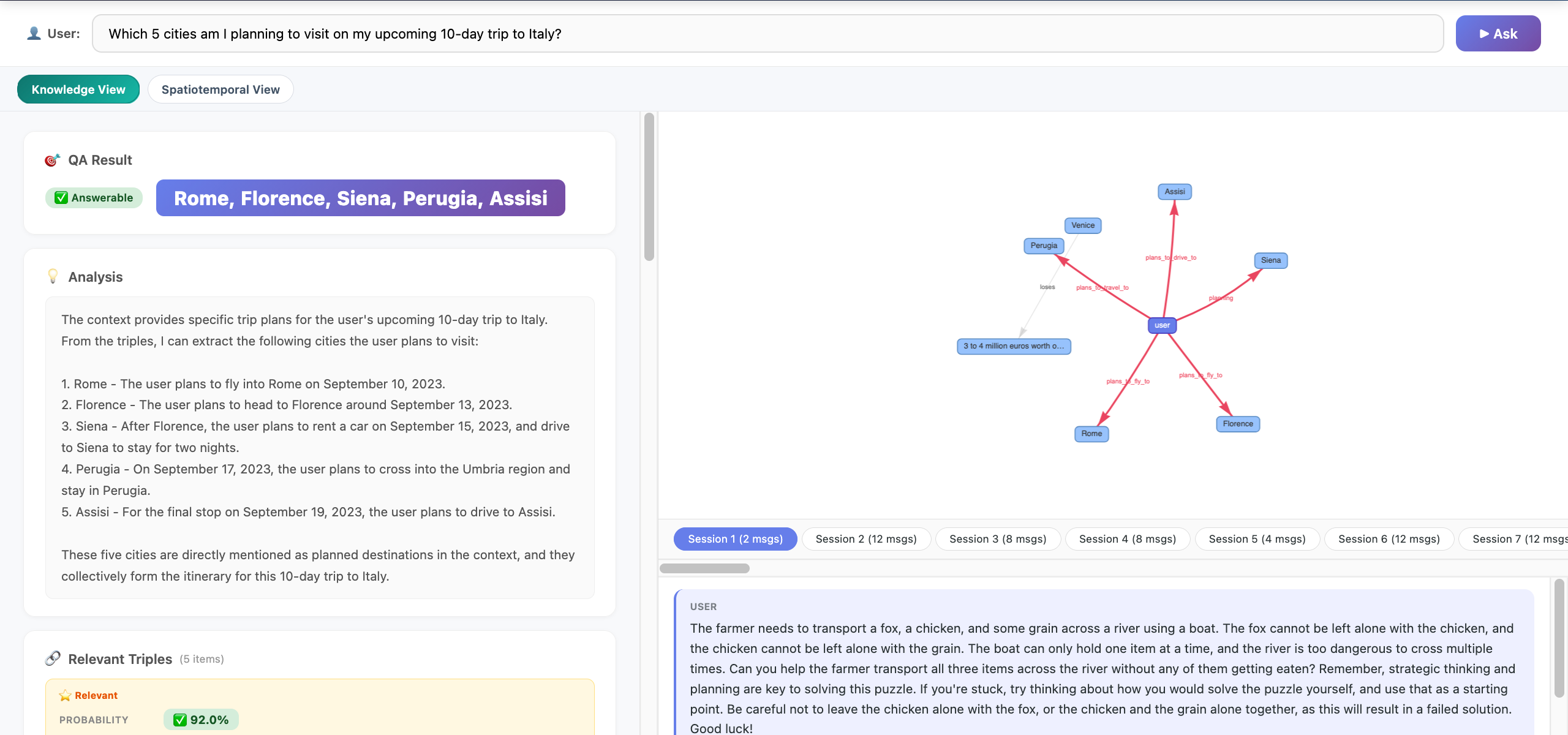}
        \end{minipage}%
    }
    \hfill
    \subfloat[Global map.\label{fig:italy_global}]{%
        \begin{minipage}[t]{0.34\textwidth}
            \centering
            \includegraphics[width=\linewidth]{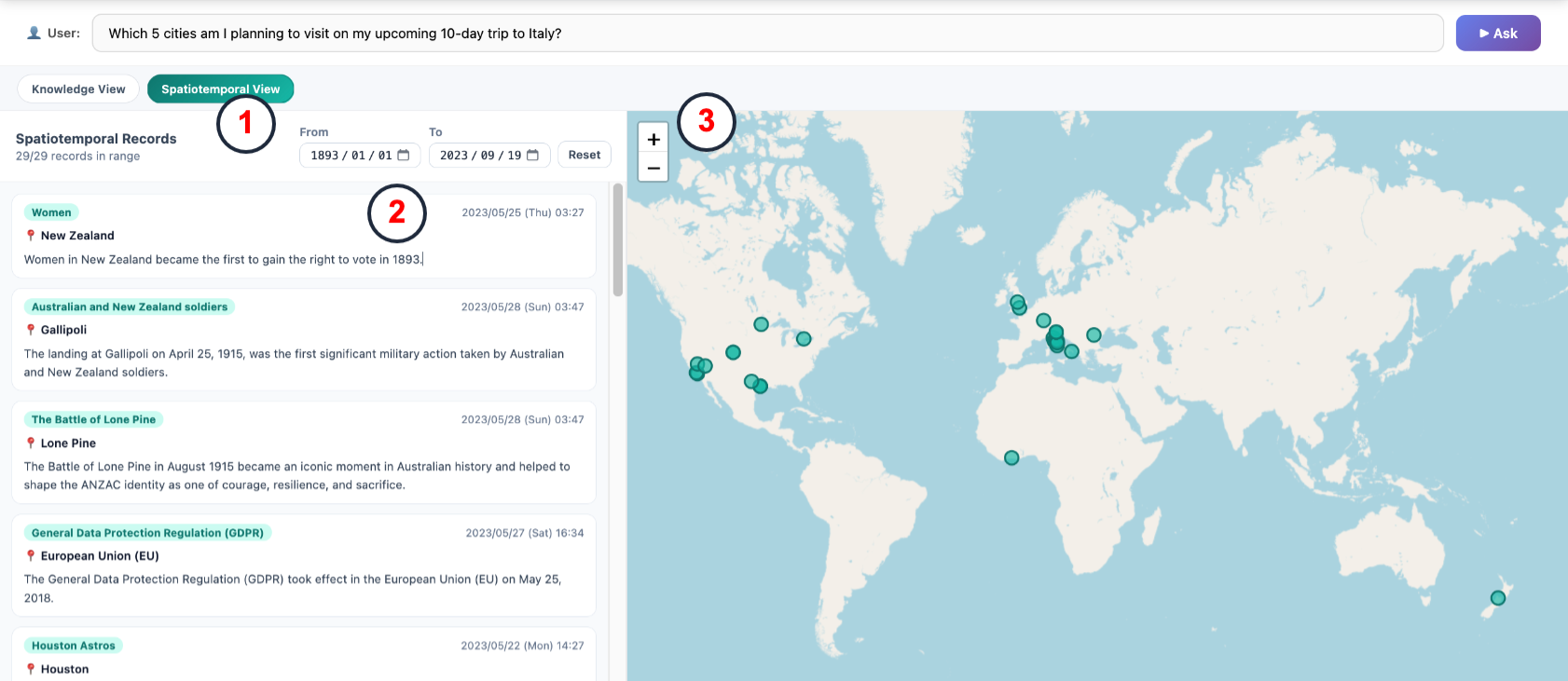}
        \end{minipage}%
    }
    \hfill
    \subfloat[Filtered map.\label{fig:italy_filtered}]{%
        \begin{minipage}[t]{0.33\textwidth}
            \centering
            \includegraphics[width=\linewidth]{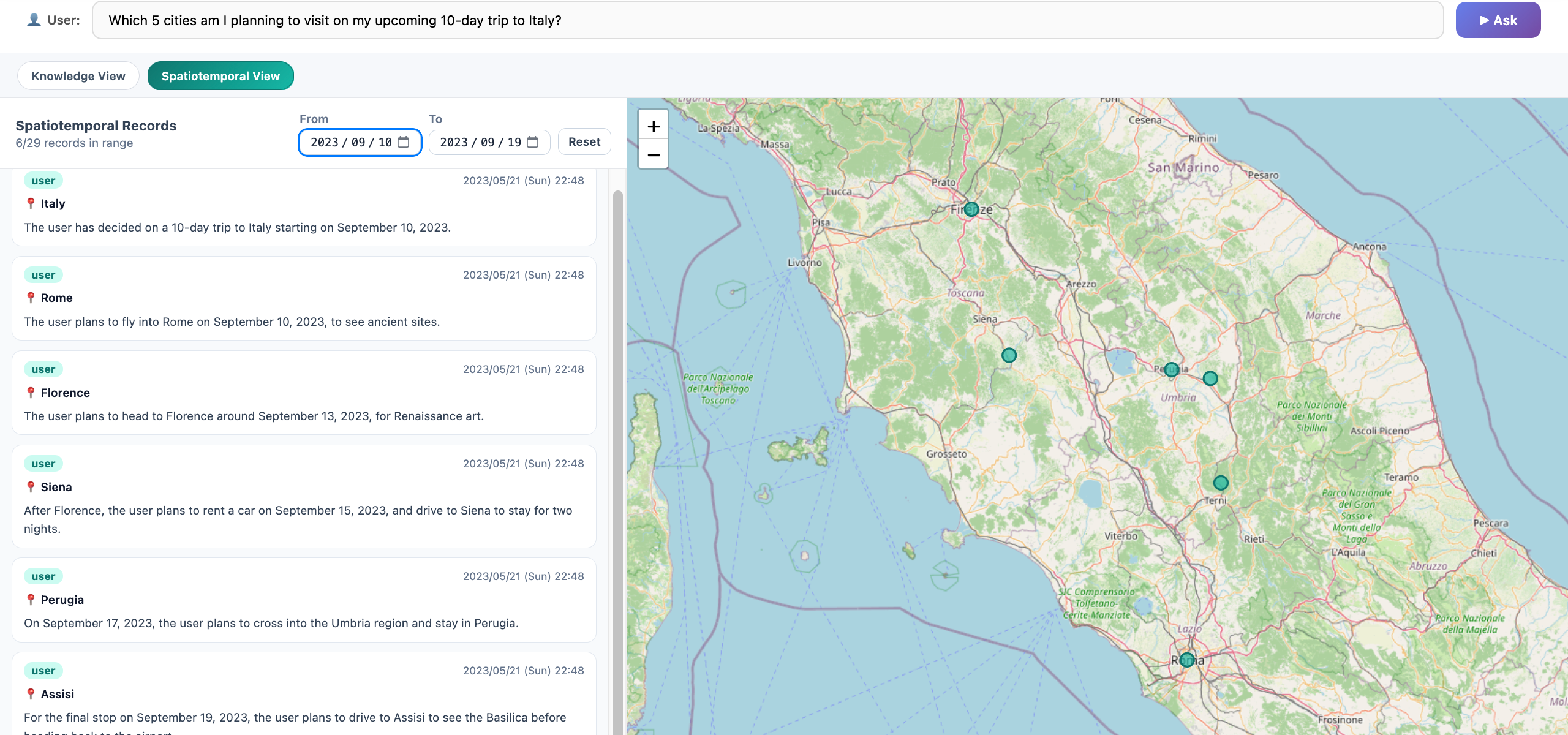}
        \end{minipage}%
    }
    \caption{spatio-temporal UI: (a) Itinerary extraction; (b) Global historical map; (c) Map filtering by dynamic time window.}
    \label{fig:spatio-temporal_demo}
\end{figure*}

PolyMemDB's backend relies on FastAPI and Pydantic-AI, while its frontend provides an interactive visual dashboard. At the conference demonstration, attendees will experience how PolyMemDB extracts and infers information from conversations spanning long periods.

\noindent\textbf{Scenario 1: Long-Session QA with Fine-Grained Provenance.} As shown in~\cref{fig:demo_main}, we introduce a long-session case (ID: d905b33f) from the LongMemEval~\cite{wuLongMemEvalBenchmarkingChat2024} benchmark. This conversation spans 48 sessions; following ingestion by PolyMemDB, a memory graph containing 229 entities and 221 relationships was successfully constructed. Through the interactive analysis interface, attendees can track the system's complete process of answering a question: First is the \textcircled{1} Query Input (e.g., entering a question involving long-range dependencies: \textit{``What percentage discount did I get on the book by my favorite author?''}), followed by the completion of \textcircled{2} Answer Generation (the retrieval outputs \textit{20\%}). Subsequently, through \textcircled{3} Reasoning Chain Parsing, the interface shows attendees how the \ac{llm} arrived at the final answer. To demonstrate that the answer is not an \ac{llm} hallucination, the system lists all relevant triplets supporting this answer, the observational facts attached to the edges, and the positive drive probability calculated based on temporal decay within the \textcircled{4} Evidence panel. Finally, the interface \textcircled{5} shows the core relational subgraph of the matched entities to reduce visual clutter, while \textcircled{6} presents the raw corpus.

\noindent\textbf{Scenario 2: Spatio-Temporal Memory Evolution.} To demonstrate PolyMemDB's capabilities in handling complex spatio-temporal constraints and long-term context summarization, we designed a second demonstration scenario (based on the extended LongMemEval~\cite{wuLongMemEvalBenchmarkingChat2024} case ID: e47becba). \textit{When the user asks the system to summarize the ``graduation trip'' itinerary from a lengthy historical dialogue}, PolyMemDB not only automatically extracts and generates a structured itinerary response (as shown in~\cref{fig:italy_knowledge}), but also provides an interactive analysis dashboard to visualize the spatio-temporal evolution of heterogeneous memories intuitively: First, through the Global Spatio-Temporal Map, the system panel lists all geographical locations and associated facts mentioned in the long conversation (\cref{fig:italy_global});
subsequently, attendees can operate the interface's \textcircled{1} Dynamic Time Selector to set a specific time window. When limited to the ``graduation trip planning period'', the system automatically filters out irrelevant locations, synchronously renders a \textcircled{2} Fine-grained Fact List in the panel to trace the underlying evidence sources of these spatio-temporal memories, and precisely highlights the 5 Italian cities the user plans to visit on the \textcircled{3} Geospatial Map (\cref{fig:italy_filtered}).

\begin{figure}[htbp]
    \centering
    \includegraphics[width=0.95\columnwidth]{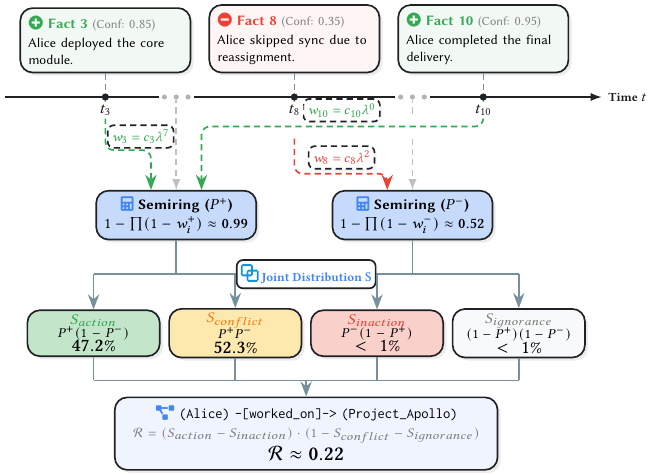}
    \caption{Probabilistic inference pipeline. Decayed historical facts are aggregated into a 4D distribution and synthesized into a net reliability score to resolve factual conflicts.}
    \label{fig:prob_inference}
\end{figure}

\noindent\textbf{Scenario 3: Dynamic Probabilistic Inference for Factual Conflicts.} We ask the question ``\textit{Was Alice actively contributing to Project Apollo in September 2024?}'' across 10 months of history (simplified in \cref{fig:prob_inference}). Using temporal decay ($\lambda=0.8$) and semiring aggregation, the engine computes a 4D cognitive distribution for her overall project involvement. Although Alice's recent triumphant delivery drives a strong positive state ($S_{action}=47.2\%$), historical setbacks like client budget freezes and a Q3 emergency reassignment trigger a dominant conflict state ($S_{conflict}=52.3\%$), penalizing the overall net reliability ($\mathcal{R} \approx 0.22$). Guided by these metrics, the system identifies her project involvement as a discontinuous event rather than a uniform historical fact. The low reliability score instructs the \ac{llm} to prioritize localized temporal evidence over the aggregated final outcome. When evaluating the requested timeframe, the engine isolates the specific records from September via the evidence chain. The underlying provenance explicitly indicates a negative polarity during this window, noting her temporary reassignment to another account. Consequently, the system answers ``no'' for that specific month, accurately contextualizing this localized absence within her overall successful delivery of the project.

\bibliographystyle{IEEEtran}
\bibliography{references}

\end{document}